\documentclass[journal]{IEEEtran}

\usepackage{pgfplots}
\usepackage[bookmarks,colorlinks]{hyperref}
\usepackage[linesnumbered,ruled,lined]{algorithm2e}
\usepackage{enumitem}
\usepackage{algpseudocode}
\usetikzlibrary{shapes.multipart,intersections}
\usepackage{cite}
\usepackage{amsmath,amssymb,amsfonts,amsthm,steinmetz}
\usepackage{graphicx}
\usepackage{mathrsfs}  
\usepackage{textcomp}
\usepackage{acronym}
\usepackage{xcolor}
\usepackage{upgreek,xspace}

\usepackage{tikz}
\usetikzlibrary{calc}
\makeatletter
\newcommand{\gettikzxy}[3]{%
  \tikz@scan@one@point\pgfutil@firstofone#1\relax
  \edef#2{\the\pgf@x}%
  \edef#3{\the\pgf@y}%
}
\makeatother

\usepackage[draft]{todonotes}

\usepackage{esvect}

\usepackage{times}
\usepackage{bm}
\usepackage{amsmath}
\usepackage{amssymb}
\usepackage{stmaryrd}
\usepackage{babel}
\usepackage{graphics, graphicx}
\usepackage{xcolor}
\usepackage{gensymb}
\usepackage{cite}
\usepackage{enumitem}
\usepackage{url}

\begin{document}

\title{Systematic Physics-Compliant Analysis \\of Over-the-Air Channel Equalization \\in RIS-Parametrized Wireless Networks-on-Chip}

\author{Jean~Tapie,~Hugo~Prod'homme,~Mohammadreza~F.~Imani,~\IEEEmembership{Member,~IEEE},~and~Philipp~del~Hougne,~\IEEEmembership{Member,~IEEE}
\thanks{
J.~Tapie, H.~Prod'homme, and P.~del~Hougne are with Univ Rennes, CNRS, IETR - UMR 6164, F-35000, Rennes, France (e-mail: \{jean.tapie; hugo.prodhomme; philipp.del-hougne\}@univ-rennes.fr).
}
\thanks{
M.~F.~Imani is with the School of Electrical, Computer, and Energy Engineering, Arizona State University, Tempe, AZ 85287, USA (e-mail: mohammadreza.imani@asu.edu).
}
\thanks{J.~Tapie and H.~Prod'homme contributed equally to this work.}
\thanks{\textit{(Corresponding Author: Philipp del Hougne.)}}
}

\maketitle

\begin{abstract}
Wireless networks-on-chip (WNoCs) are an enticing complementary interconnect technology for multi-core chips but face severe resource constraints. Being limited to simple on-off-keying modulation, the reverberant nature of the chip enclosure imposes limits on allowed modulation speeds in sight of inter-symbol interference, casting doubts on the competitiveness of WNoCs as interconnect technology. Fortunately, this vexing problem was recently overcome by parametrizing the on-chip radio environment with a reconfigurable intelligent surface (RIS). By suitably configuring the RIS, selected channel impulse responses (CIRs) can be tuned to be (almost) pulse-like despite rich scattering thanks to judiciously tailored multi-bounce path interferences. However, the exploration of this ``over-the-air'' (OTA) equalization is thwarted by (i) the overwhelming complexity of the propagation environment, and (ii) the non-linear dependence of the CIR on the RIS configuration, requiring a costly and lengthy full-wave simulation for every optimization step. Here, we show that a reduced-basis physics-compliant model for RIS-parametrized WNoCs can be calibrated with a single full-wave simulation. Thereby, we unlock the possibility of predicting the CIR for any RIS configuration almost instantaneously without any additional full-wave simulation. We leverage this new tool to systematically explore OTA equalization in RIS-parametrized WNoCs regarding the optimal choice of delay time for the RIS-shaped CIR's peak. We also study the simultaneous optimization of multiple on-chip wireless links for broadcasting. Looking forward, the introduced tools will enable the efficient exploration of various types of OTA analog computing in RIS-parametrized WNoCs.
\end{abstract}


\section{Introduction}\label{sec_Introduction}

Multicore chips, integrating a multitude of independent cores within a single chip, increasingly struggle to guarantee low-latency information exchange between far apart cores, implying that communication increasingly becomes the bottleneck for their computing performance~\cite{nychis2012chip,karkar2016survey}. Communication between cores is currently based on packet-switched networks composed of routers and wires~\cite{wentzlaff2007chip}. This wired interconnect technology experiences unfavorable scalings as the number of cores increases, mainly due to poor latency on multi-hop trajectories as well as significant Ohmic losses~\cite{marculescu2008outstanding,todri}. Moreover, their fixed hardware lacks traffic-adaptive flexibility. These considerations spur efforts to explore alternative interconnect technologies that can complement wired interconnects for communication between far-apart cores~\cite{franques2021widir}.

On the one hand, a wide range of guided network-on-chip (NoC) technologies is currently being explored in the radio-frequency, nanophotonic and plasmonic regimes. Although some of these promise significant benefits in terms of energy efficiency and bandwidth~\cite{miller2009device,dionne2010silicon,sun2015single,wade2020teraphy}, the required technologies (e.g., integrated light sources) are not yet mature and remain costly; morever, similar to wired interconnects, these NoCs rely on inflexible waveguide networks that may require multi-router hops and lack the desired flexibility to adapt to traffic needs. On the other hand, wireless NoCs (WNoCs) rely on mature technology, naturally offer the sought-after flexibility, and naturally bypass the issue of multi-router hops, thereby promising low-latency broadcasting~\cite{chang2001rf,ganguly2010scalable,matolak2012wireless,deb,6814853,shamim2016wireless,abadal2022graphene,calo2022reconfigurable}.
However, the envisioned WNoCs are highly resource-constrained networks with very limited modulation and processing capabilities~\cite{laha2014new,6814853,markish2015chip,abadal2019wave}. WNoCs are typically based on on-off-keying (OOK), i.e., binary amplitude shift keying (BASK). 
In sight of the on-chip propagation environment, essentially a micro reverberation chamber~\cite{matolak2013channel} in which the channel impulse responses (CIRs) have many taps due to the metallic chip enclosure, the OOK modulation speed must be throttled to avoid inter-symbol interference (ISI) at the receiver. Severe doubts about the competitiveness of WNoCs as complementary interconnect technology hence arise.

Various attempts at overcoming this vexing ISI concern were recently made that typically fall into one of the three following categories: 
\begin{enumerate}
    \item Using more resource-demanding modulation schemes (e.g., a time-diversity scheme~\cite{8512165} or time reversal~\cite{fatimaTR}).
    \item Controlling the level of absorption (e.g., via the thickness of the lossy silicon layer~\cite{timoneda2020engineer} or via absorbing boundaries~\cite{treguer2023broadband}) to mitigate ISI at the cost of a lower received signal strength. 
    \item Endowing the chip propagation environment with programmability~\cite{imani2021smart} to equalize wireless channels ``over the air'' (OTA). Specifically, by integrating a reconfigurable intelligent surface (RIS) into the chip package, the on-chip radio environment becomes to some extent tunable. Thereby, the interference of the multitude of multi-bounce paths linking the transmitter to the receiver can be tailored, and an (almost) pulse-like CIR can be imposed between a selected pair of antennas despite the rich scattering.
\end{enumerate}
The third approach holds the promise of being able to use the very frugal OOK modulation without ISI and without signal strength penalty, provided an on-chip RIS can be integrated and suitably configured.
Although Ref.~\cite{imani2021smart} demonstrated the feasibility of such RIS-assisted over-the-air (OTA) channel equalization for WNoCs in full-wave simulations, a systematic analysis of the concept remained out of reach due to the prohibitive cost of the underlying full-wave simulations. On the one hand, the chip is electrically very large, making every single simulation very costly. On the other hand, the mapping from RIS configuration to CIR is non-linear due to proximity-induced and reverberation-induced coupling between the RIS elements~\cite{rabault2023tacit}, requiring an iterative optimization of the RIS configuration with one new full-wave simulation per iteration. 

In the present paper, we demonstrate that a compact physics-compliant model of the RIS-parametrized CIRs can be calibrated to any specific chip architecture of interest based on a single full-wave simulation. Recently, there have been multiple efforts to conceive physics-compliant models for RIS-parametrized radio environments (not at the chip scale, but the underlying physics is scale-invariant). These describe the RIS elements either in terms of tunable load impedances~\cite{gradoni_EndtoEnd_2020,shen2021modeling,badheka2023accurate,akrout2023physically,mursia2023saris} or tunable polarizabilities~\cite{PhysFad,prod2023efficient}, and -- except for Refs.~\cite{PhysFad,prod2023efficient,mursia2023saris} -- are limited to a simple free-space propagation environment. However, even Refs.~\cite{PhysFad,prod2023efficient,mursia2023saris} are limited to abstract scattering environments composed of discrete dipoles surrounded by free space. By contrast, the chip environment involves continuous metallic walls and various dielectric layers (see Sec.~\ref{sec_wnoc} below) for which a tractable discrete-dipole description is not obvious. Fortunately, the effect of the scattering environment can be understood as merely contributing additional long-range reverberation-induced coupling effects between the primary wireless entities of interest (antennas and RIS elements)~\cite{prod2023efficient}. Mathematically, this follows straight-forwardly from the block matrix inversion lemma and yields an equivalent representation of the system in the reduced basis of primary wireless entities as opposed to the conventional canonical basis~\cite{prod2023efficient}. This reduced-basis representation is not limited to discrete scattering objects and has recently been experimentally validated in 3D for an irregularly shaped scattering enclosure with continuous metallic walls of unknown geometry and material composition~\cite{sol2023experimentally}. In Ref.~\cite{sol2023experimentally}, a gradient-descent optimization was used to identify the parameters of a reduced-basis tunable-polarizability-based physical model from experimentally obtained calibration data. In the present paper dedicated to the study of WNoCs with full-wave simulations, we show that the parameters of a reduced-basis tunable-load-impedance-based physical model can be directly obtained with a single full-wave simulation.

Once calibrated, we can use the physical model of the RIS-parametrized WNoC to correctly and efficiently predict (almost instantaneously and without any additional full-wave simulation) the CIR for any arbitrary RIS configuration -- despite the complicated non-linear mapping from RIS configuration to CIR as well as the overwhelming complexity of the on-chip radio environment. Thereby, we unlock the possibility of systematically studying RIS-parametrized WNoCs. We leverage this new tool to systematically explore the RIS-based CIR optimization for OTA equalization. Indeed, the preliminary physical-model-based studies of OTA equalization in Refs.~\cite{PhysFad,hugo_eucap} suggested that the optimal choice of delay time for the optimized CIR's peak appears to depend on the amount of reverberation as well as on setting-specific features. The reason relates to the line-of-sight (LOS) tap of the CIR that originates from the LOS path which did not encounter the RIS and is hence non tunable. In the regime of moderate and weak reverberation, the LOS tap plays a significant role in the CIR and hence the best strategy may be to aim to suppress all taps other than the LOS tap. In contrast, in the regime of strong reverberation, it is possible to strongly enhance taps at later delay times such that the LOS tap becomes negligible in comparison. Given our efficient physical-model-based WNoC simulator, we systematically explore these regimes in the WNoC context in the present paper. Moreover, we consider the simultaneous optimization of multiple CIRs for broadcast scenarios.

This paper is organized as follows. In Sec.~\ref{sec_Theory}, we establish the theory of our reduced-basis tunable-load-impedance-based physical model for RIS-parametrized WNoCs. In Sec.~\ref{sec_wnoc}, we describe the considered prototypical RIS-parametrized WNoC and characterize the on-chip RIS element. In Sec.~\ref{sec_valid}, we validate in full-wave simulation the physical model developed in Sec.~\ref{sec_Theory}.
In Sec.~\ref{syststud}, we present a systematic study of RIS optimization for OTA equalization in WNoCs, identifying the optimal choice of the delay time for the optimized CIR's peak. In Sec.~\ref{sec_multi}, we explore the simultaneous OTA equalization of multiple CIRs for broadcast scenarios. We close in Sec.~\ref{sec_Conclusion} with a conclusion and outlook to future work.

\textit{Notation.} $\mathbf{A} = \mathrm{diag}(\mathbf{a})$ denotes the diagonal matrix $\mathbf{A}$ whose diagonal entries are $\mathbf{a}$. 
$\left[ \mathbf{A}^{-1} \right]_\mathcal{BC}$ denotes the block of $\mathbf{A}^{-1}$ selected by the sets of indices $\mathcal{B}$ and $\mathcal{C}$. 
$\mathbf{I}_a$ denotes the $a \times a$ identity matrix.
$\mathbf{0}_\mathcal{B}$ [$\mathbf{0}_\mathcal{BC}$] denotes the $b\times 1$ vector [$b\times c$ matrix] whose entries are all zero, where $b$ and $c$ are the number of indices included in $\mathcal{B}$ and $\mathcal{C}$, respectively.

\section{Theory}\label{sec_Theory}

The wireless entities of primary interest (antennas and RIS elements) are naturally discrete. By contrast, the scattering environment in WNoCs is composed of continuous scattering objects (metallic walls and dielectric slabs). We consider in this theory section a generic scenario involving $N_\mathrm{T}$ transmitting antennas, $N_\mathrm{R}$ receiving antennas, and $N_\mathrm{S}$ RIS elements. There are hence $N=N_\mathrm{A}+N_\mathrm{S}$ primary wireless entities, where $N_\mathrm{A} = N_\mathrm{T}+N_\mathrm{R}$. Let us denote by $\mathcal{T}$, $\mathcal{R}$ and $\mathcal{S}$ the sets of $N_\mathrm{T}$, $N_\mathrm{R}$ and $N_\mathrm{S}$ indices assigned to the transmitters, receivers and RIS elements, respectively. We further define $\mathcal{A} = \mathcal{T} \cup \mathcal{R}$ and $\mathcal{N} = \mathcal{A}  \cup \mathcal{S}$.

The frequency-dependent wireless channel matrix $\mathbf{H}(f)\in\mathbb{C}^{N_{\mathrm{R}}\times N_{\mathrm{T}}}$ is the $\mathcal{RT}$ block of the system's scattering matrix $\mathbf{S}(f)\in\mathbb{C}^{N_{\mathrm{A}}\times N_{\mathrm{A}}}$ that is defined in an obvious manner by the $N_{\mathrm{A}}$ antennas:
\begin{equation}
    \mathbf{H}(f) = \left[ \mathbf{S}(f)\right]_\mathcal{RT},
    \label{relationHS}
\end{equation}
where
\begin{equation}
    \mathbf{S}(f) = \begin{bmatrix} 
	\mathbf{R^{in}}(f)  & \mathbf{H}^T(f) \\	\mathbf{H}(f)  & \mathbf{R^{out}}(f)
 \end{bmatrix}
 \label{eqS},
\end{equation}
under the assumption of reciprocity, and $\mathbf{R^{in}}(f)\in \mathbb{C}^{N_\mathrm{T} \times N_\mathrm{T}}$ is the reflection matrix capturing the waves scattered back into the transmission lines attached to the transmitting antennas.

Equivalently, our $N_{\mathrm{A}}$-port system can be represented by an impedance matrix $\mathbf{Z}(f)\in\mathbb{C}^{N_{\mathrm{A}}\times N_{\mathrm{A}}}$ that is related to $\mathbf{S}(f)$ as follows:
\begin{equation}
    \mathbf{S}(f) = \left( \mathbf{Z}(f) + Z_0 \mathbf{I}_{N_\mathrm{A}}\right)^{-1} \left( \mathbf{Z}(f) - Z_0 \mathbf{I}_{N_\mathrm{A}}\right),
    \label{relationSZ}
\end{equation}
where $Z_0$ is the characteristic impedance of the transmission lines attached to the antennas (typically $50 \ \Omega$). Expressing $\mathbf{H}(f)$ in terms of $\mathbf{Z}(f)$ is more cumbersome than in terms of $\mathbf{S}(f)$ but allows us to formalize the dependence of $\mathbf{H}(f)$ on the RIS configuration in the following. 

\textit{Remark:} Throughout this paper, we do \textit{not} make the widespread ``unilateral'' approximation proposed in Ref.~\cite{ivrlavc2010toward} that would consist in assuming $\left[ \mathbf{Z} \right]_\mathcal{TR} = \mathbf{0}$ in order to simplify the expression for how $\mathbf{H}(f)$ depends on $\mathbf{Z}(f)$ at the cost of formally breaking reciprocity. 

The $N_\mathrm{S}$ RIS elements are now treated as $N_\mathrm{S}$ auxiliary ports which are not connected to transmission lines through which waves can enter and exit the system but instead they are terminated by tunable load impedances. This gives rise to an augmented $N$-port auxiliary impedance matrix $\hat{\mathbf{Z}}(f) \in\mathbb{C}^{N\times N}$ that is related to the measurable $\mathbf{Z}(f)$ as follows:
\begin{equation}
    \mathbf{Z}(f) = \hat{\mathbf{Z}}_\mathcal{AA}(f) - \hat{\mathbf{Z}}_\mathcal{AS}(f) \left( \hat{\mathbf{Z}}_\mathcal{SS}(f) + \mathbf{\Phi}(f)\right)^{-1} \hat{\mathbf{Z}}_\mathcal{SA}(f),
    \label{Zaug}
\end{equation}
where 
\begin{equation}
    \mathbf{\Phi}(f)=\mathrm{diag} \left( \mathbf{c}(f) \right)
\end{equation}
and the RIS configuration $\mathbf{c}(f)\in \mathbb{C}^{N_\mathrm{S}\times 1}$ is a vector containing the $N_\mathrm{S}$ tunable load impedances of the RIS elements. A derivation of Eq.~(\ref{Zaug}) is provided for completeness in Appendix~\ref{AppendixA}. 

Hence, given $\mathbf{\hat{Z}}(f)$ and the allowed values that the tunable load impedances can take, $\mathbf{H}(f)$ can be worked out analytically for any desired RIS configuration $\mathbf{c}(f)$ using Eq.~(\ref{Zaug}), Eq.~(\ref{relationSZ}) and Eq.~(\ref{relationHS}). The generally non-linear dependence of $\mathbf{H}(f)$ on $\mathbf{c}(f)$ is apparent. By declaring the tunable lumped elements of the RIS elements as lumped ports with characteristic impedance $Z_0$, $\mathbf{\hat{Z}}(f)$ can be obtained with a full-wave solver -- irrespective of (i) the complexity of the scattering environment, (ii) whether the scattering objects are discrete or continuous, and (iii) the detailed nature of the wave propagation mechanisms (e.g., space waves, surface waves, guided waves~\cite{zhang2007propagation}). 
Thus, the tunable-load-impedance-based physical model of RIS-parametrized channels is ideally suited for the study of WNoCs based on full-wave simulations.

The fundamental assumption whose validity determines the accuracy of this approach is that the auxiliary ports are electrically small such that they have a constant current distribution. This condition is approximately satisfied by typical lumped elements like the PIN diodes considered below which are electrically small.
However, no requirement regarding the size of the RIS elements whose scattering properties are controlled by these tunable load impedances exists. As already pointed out in Ref.~\cite{konno2023generalised}, the size of the RIS element unit cell does not impact the validity of the approach. 

The CIR $h_{i,j}(t,\mathbf{c})$ between the antennas indexed $i$ and $j$ is related to the corresponding frequency-domain channel $H_{i,j}(f,\mathbf{c})$ via a Fourier transform:
\begin{equation}
    h_{i,j}(t,\mathbf{c}) = \mathcal{F}_\mathrm{R}^{-1} \{ H_{i,j}(f,\mathbf{c}) \},
\end{equation}
where $\mathcal{F}_\mathrm{R}\{\cdot\}$ denotes the Fourier transform of a real signal, and $H_{i,j}(f,\mathbf{c})$ only includes positive frequencies. The CIR envelope is then obtained as follows:
\begin{equation}
    e_{i,j}(t,\mathbf{c}) = \left| \mathcal{H} \{ h_{i,j}(t,\mathbf{c}) \} \right|,
\end{equation}
where $\mathcal{H}\{\cdot\}$ denotes the Hilbert transform.

\section{RIS-Parametrized Wireless Network-on-Chip}
\label{sec_wnoc}

\begin{figure}
\centering
\includegraphics[width=\linewidth]{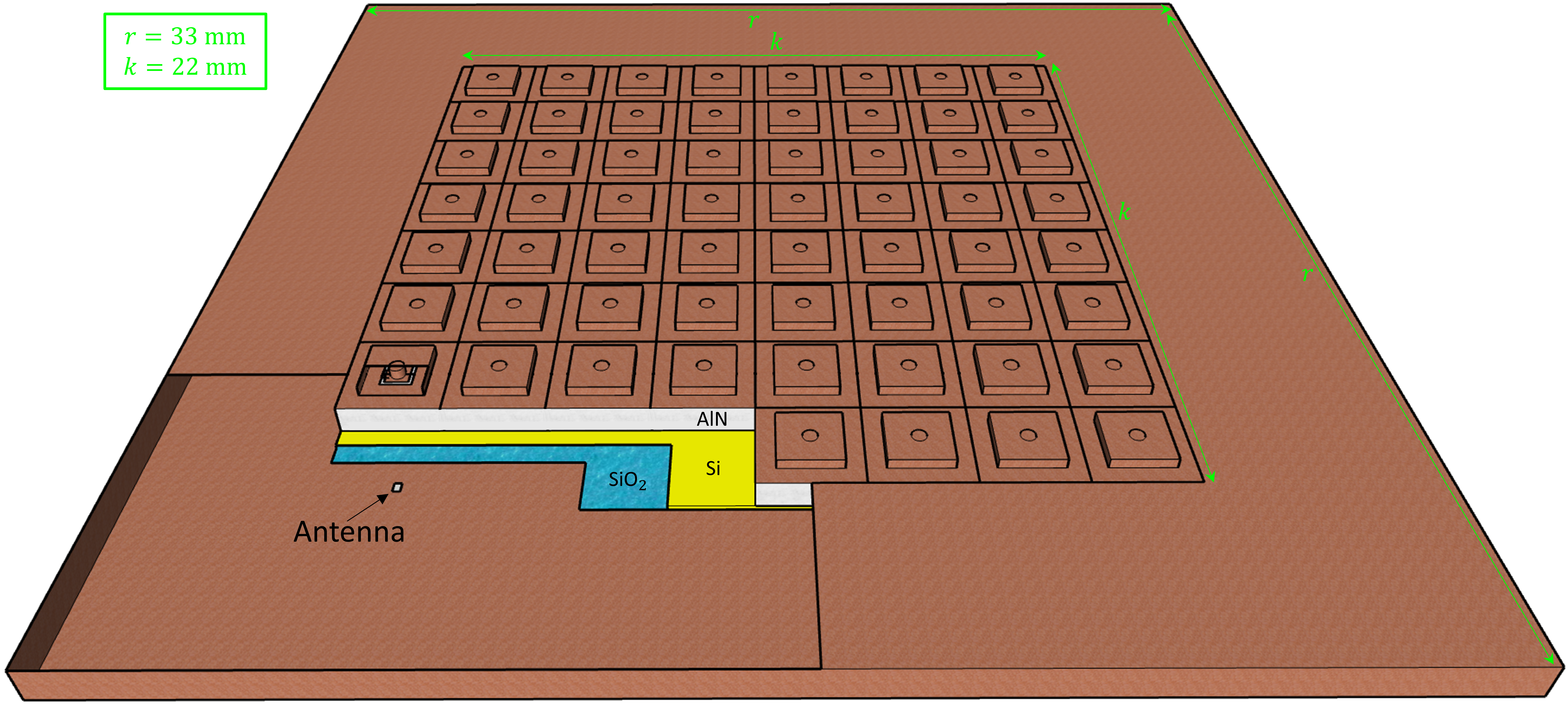}
\caption{Considered multi-core chip architecture with $8 \times 8$ on-chip RIS. Parts in the bottom left corner are partially removed to show the interior. The detailed RIS element design is shown in Fig.~\ref{fig_ris_element}. The properties of the dielectric layers are summarized in Table~\ref{table1}.}
\label{fig_chip}
\end{figure}

We consider the prototypical chip architecture shown in Fig.~\ref{fig_chip} that is simplified with respect to electromagnetically irrelevant details~\cite{imani2021smart}. The footprint of the metallic chip package is $33\times 33\ \mathrm{mm}^2$. A $22\times 22\ \mathrm{mm}^2$ large chip is centered therein and surrounded by air. The chip is modelled as a multi-layer structure composed of three dielectric slabs. On the bottom is a $0.011\ \mathrm{mm}$ thick layer of silicon dioxide ($\mathrm{SiO}_2$), on top of that is a $0.1\ \mathrm{mm}$ thick layer of silicon ($\mathrm{Si}$), and thereupon is a $0.8\ \mathrm{mm}$ thick layer of aluminum nitride ($\mathrm{AlN}$). 
Finite conductivity boundary conditions emulating copper surround the entire simulation domain, representing the metallic chip package on the side walls and top, and the essentially flat layer of solder bumps on the bottom.

Our WNoC operates within a $B = 10 \ \mathrm{GHz}$ wide frequency band centered on $f_0 = 60\ \mathrm{GHz}$, i.e., $55\ \mathrm{GHz}<f<65\ \mathrm{GHz}$, where $f$ denotes the frequency. We use the Ansys High Frequency Structural Simulator (HFSS) to simulate wave propagation in the chip environment using a finite-element method. 

Electrically small slots of dimensions $0.2 \times 0.3 \ \mathrm{mm}^2$ in the bottom metallic layer serve as antennas in our study. These slot dimensions are deeply sub-wavelength compared to the wavelength at 60~GHz in the materials contained within the chip -- see Table~\ref{table1}. Therefore, the radiation properties are approximately frequency flat in the considered band and the antennas have low directivity. These two features are aligned with the desire for broadband and broadcast operation in WNoCs.\footnote{Future work can refine these antennas based on existing literature about on-chip antennas~\cite{markish2015chip,rayess2017antennas,narde2019intra,narde2020antenna,treguer2023broadband}. Note that this literature so far often deals with unpackaged chip configurations in which the ISI problem that is encountered in typical fully packed chips is much weaker.}  Note that in our considered scenario there is essentially no LOS path with a delay related to the spatial separation between antenna pairs. Instead, the shortest path between any two antennas involves one bounce of the metallic ceiling of the chip package.

\begin{table} [t]
\caption{Summary of relative permittivity and wavelength \\at 60~GHz for the considered materials.}
\label{tbl:Symbols}
\begin{center}
\begin{tabular}{ |c|c|c|c| } 
\hline
Material & Relative & Wavelength  \\
~ & Permittivity & at 60~GHz \\
\hline
Air & $1$ & $5.0 \ \mathrm{mm}$ \\ 
$\mathrm{SiO}_2$ & $4$ & $2.5 \ \mathrm{mm}$ \\ 
$\mathrm{Si}$ & $11.9$ & $1.4 \ \mathrm{mm}$ \\ 
$\mathrm{AlN}$ & $8.8$ & $1.7 \ \mathrm{mm}$ \\ 
\hline
\end{tabular}
\end{center}
\label{table1}
\end{table}

\begin{figure}[b]
\centering
\includegraphics[width=\linewidth]{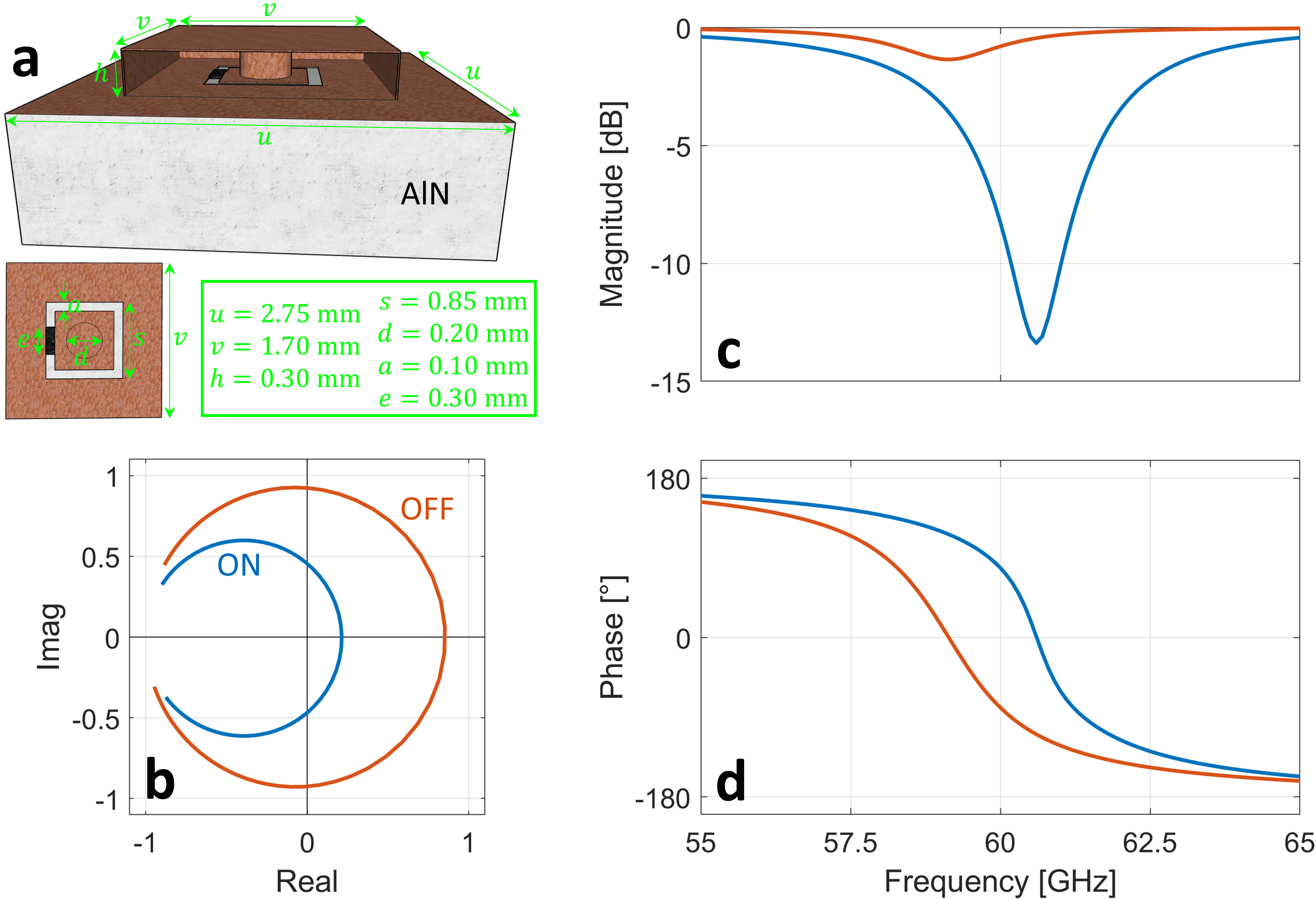}
\caption{RIS element design and characterization. a) RIS element design. One metallic wall is removed to show the interior. The black block is the lumped element with tunable impedance. b) Normal-incidence reflection coefficient of the unit cell from (a) as a function of frequency and for the two considered load impedance values, plotted in the complex frequency plane. c,d) Representation of the data from (b) in terms of magnitude (c) and phase (d) as a function of frequency.}
\label{fig_ris_element}
\end{figure}

The metallic ceiling of the chip package is patterned on its inside with $8\times 8$ regularly spaced RIS elements, as shown in Fig.~\ref{fig_chip}. The RIS element design is that proposed as ``alternative on-chip RIS design'' in Fig.~7 of Ref.~\cite{imani2021smart} and summarized in Fig.~\ref{fig_ris_element}a. We consider here that the tunable load impedance is implemented by a PIN diode such as the solderable, flip-chip Aluminum Gallium Arsenide (AlGaAs) PIN diode MADP-000907-14020x which is useable up to 70~GHz~\cite{macom}. Based on its data sheet, we consider that the two possible choices of load impedance are $\eta^\mathrm{ON}(f) = 5.2~\Omega$ and $\eta^\mathrm{OFF}(f) = (\jmath\omega C)^{-1}$, where $\jmath=\sqrt{-1}$, $\omega = 2\pi f$ and $C=25 \ \mathrm{fF}$. The in-plane dimensions of the RIS element are comparable to the wavelength at 60~GHz in AlN (while the third dimension is small compared to the wavelength). We deliberately use such a large RIS element because it limits the number of tunable lumped elements (each of which requires its own DC bias voltage in practice). Note that our use case for RIS in a rich-scattering environment does not have the same requirements as the commonly studied use case involving RIS in free space for which oftentimes sub-wavelength RIS elements are considered.

In Fig.~\ref{fig_ris_element} we characterize the RIS element under normal incidence, working with periodic boundary conditions on all four sides, i.e., assuming an infinite array of identically configured RIS elements. The simulation domain size (denoted by $u$ in Fig.~\ref{fig_ris_element}a) is chosen such that the spacing of the RIS elements of the infinite array is the same as in the chip environment in Fig.~\ref{fig_chip}. The phase difference of the response for the two possible configurations reaches its maximum of $160^\circ$ at 60~GHz, i.e., in the middle of the considered frequency band. The amplitude response of the RIS element is intertwined with its phase response, as expected for a Lorentzian resonator. While the magnitude only has a very weak dip of $-1.4 \ \mathrm{dB}$ at 59.1~GHz in the OFF state, it has much stronger dip of $-13.4 \ \mathrm{dB}$ at 60.6~GHz in the ON state. Switching between ON and OFF states thus affects both the phase and the magnitude of the scattered wave.

\begin{figure}[h]
\centering
\includegraphics[width=0.9\linewidth]{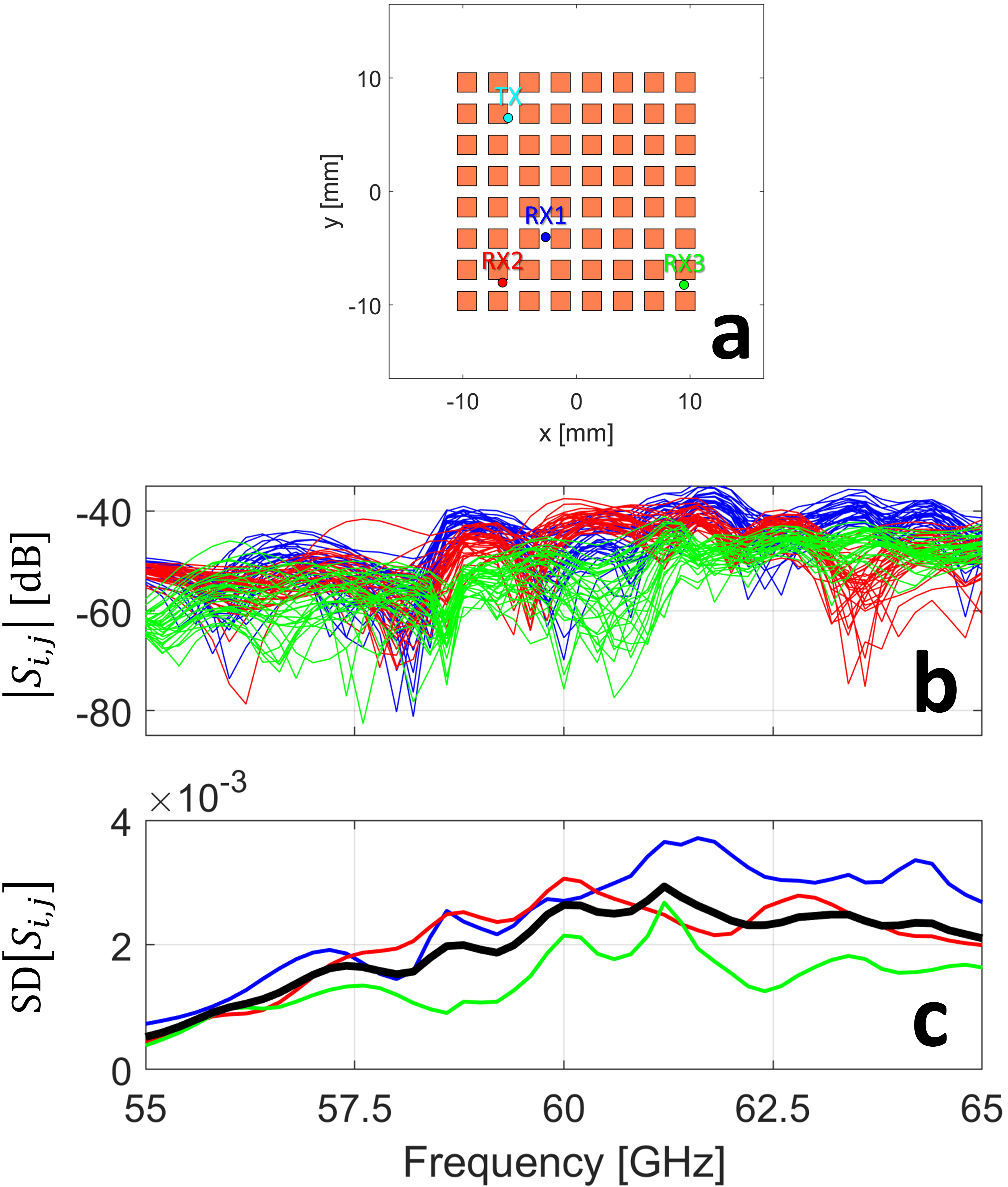}
\caption{Fluctuations of three selected on-chip wireless channels over random RIS configurations. a) Selected locations for the transmitter and the three receivers. b) 100 random realizations of the three considered channels (same color code as in a). c) Standard deviation of the wireless channels across 100 random configurations. $\mathrm{SD}[\cdot]$ denotes the standard deviation.}
\label{fig_in_situ_charac}
\end{figure}

We also characterize our on-chip RIS directly in situ in terms of how the wireless channels fluctuate for random RIS configurations. In our use case, waves from all possible angles are incident on the RIS due to the rich scattering inside the chip enclosure. We focus on a broadcast scenario with one transmitter in the bottom left part of the chip and the receivers on the right side of the chip, as shown in Fig.~\ref{fig_in_situ_charac}a. The magnitudes of these three wireless channels are traced for 100 random RIS configurations in Fig.~\ref{fig_in_situ_charac}b and reveal that the RIS significantly impacts the wireless channel across the entire considered frequency band. This is further corroborated in terms of the standard deviation of the complex-valued wireless channel across the 100 random RIS configurations in Fig.~\ref{fig_in_situ_charac}c.

\section{Full-Wave Physical Model Validation}
\label{sec_valid}

In our full-wave simulation, we can treat the tunable-impedance elements either as lumped elements (i.e., imposing a specific load impedance on each element) in which case we can extract the $N_\mathrm{A} \times N_\mathrm{A}$ matrices $\mathbf{S}(f,\mathbf{c})$ and $\mathbf{Z}(f,\mathbf{c})$ that correspond to the chosen RIS configuration $\mathbf{c}(f)$; or we can treat the tunable elements as lumped ports (with $Z_0=50\ \Omega$ characteristic impedance) in which case we can extract the $N \times N$ augmented matrices $\mathbf{\hat{S}}(f)$ and $\mathbf{\hat{Z}}(f)$ which are independent of $\mathbf{c}$. In this section, we validate the approach developed in Sec.~\ref{sec_Theory} for deriving $\mathbf{H}(f,\mathbf{c})$ from $\mathbf{\hat{Z}}(f)$ by comparing it for specific configurations $\mathbf{c}(f)$ with the corresponding directly extracted $\mathbf{H}(f,\mathbf{c}) = \left[ \mathbf{S}(f,\mathbf{c})\right]_\mathcal{RT}$.

\begin{figure}[hhh]
\centering
\includegraphics[width=\linewidth]{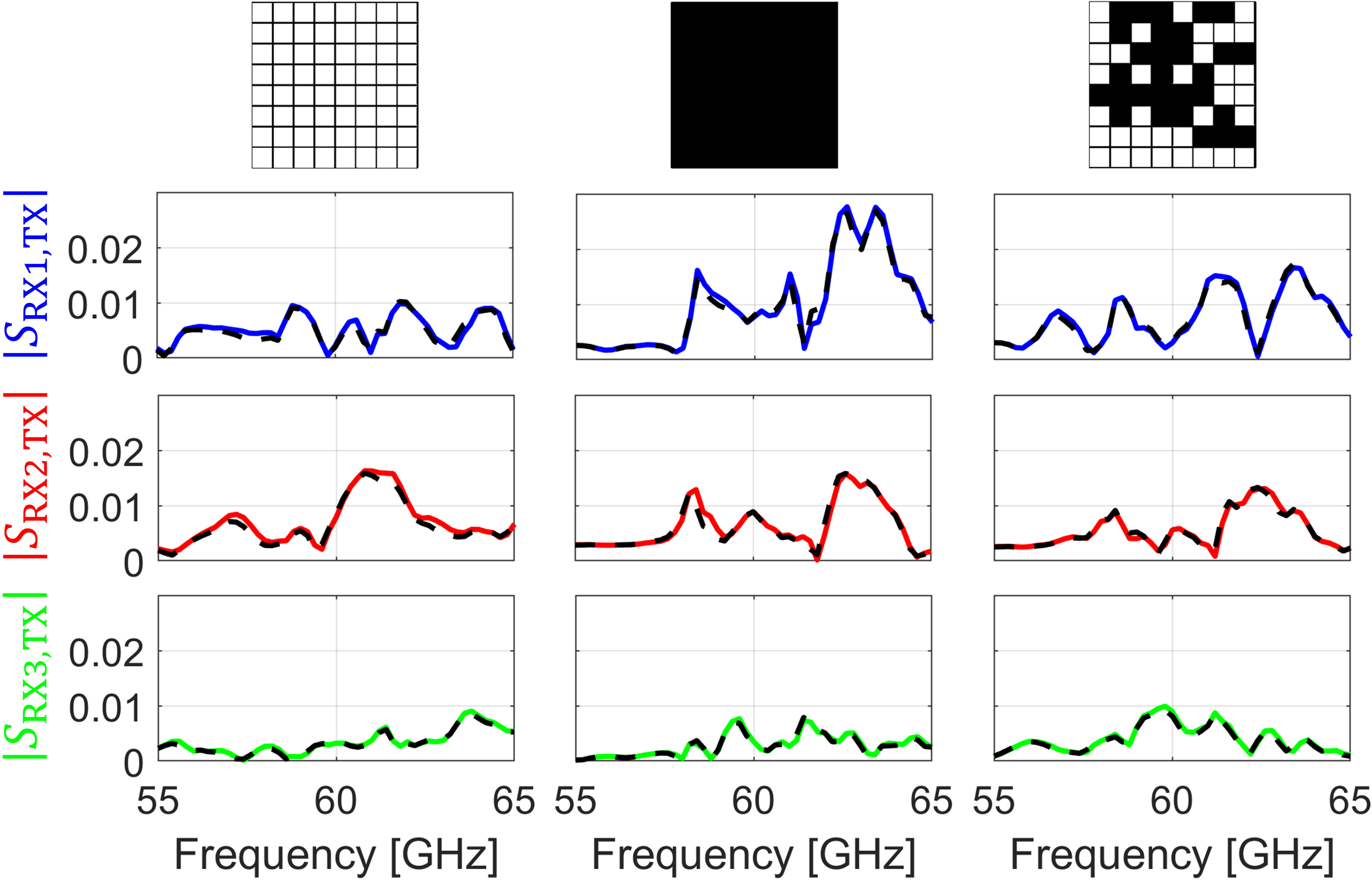}
\caption{Validation of the physics-compliant model from Sec.~\ref{sec_Theory} in the setting from Sec.~\ref{sec_wnoc}. For the three considered wireless channels (color-coded, one in each row), the model's prediction (continuous colored line) is compared to the ground truth (black dashed line) in terms of the channel magnitudes, for the three RIS configurations indicated on top (white = ON; black = OFF).}
\label{fig_valid}
\end{figure}

In Fig.~\ref{fig_valid} we show for each of the three wireless channels of interest, and for three RIS configurations in each case (all ON, all OFF, and a random configuration), the comparison between ground truth (simulation with lumped elements) and predicted (using method from Sec.~\ref{sec_Theory}) wireless channel magnitudes. Overall, satisfactory agreement is seen in all cases. Broadband agreement in terms of amplitude as seen in Fig.~\ref{fig_valid} implies also agreement in terms of phase due to the Kramers-Kroning relation. In addition, we provide additional time-domain validation in Fig.~\ref{fig_results2} for the optimized RIS configurations. 
We attribute minor differences in Fig.~\ref{fig_valid} and Fig.~\ref{fig_results2} to (i) differences in the mesh utilized in the different simulations, and (ii) the finite port size that can be considered as infinitesimally small only in first approximation.

\bigskip

\section{Systematic Study of OTA Equalization \\ of a Single CIR}
\label{syststud}

Having formulated, calibrated and validated our physics-compliant model for RIS-parametrized WNoCs, we can now use it to systematically study OTA equalization in a computationally efficient manner. In this section, we begin by studying OTA equalization of a single CIR. The simultaneous equalization of multiple CIRs OTA is considered in the subsequent Sec.~\ref{sec_multi}.

The ability to tailor the temporal properties of a CIR with the spatial control offered by an RIS was experimentally demonstrated already in 2016 in the 2.5~GHz regime under rich-scattering conditions in the context of spatiotemporal focusing~\cite{del2016spatiotemporal}. In these optimizations, the objective was to maximize the CIR envelope at a given delay time. Such optimizations are not guaranteed to automatically mitigate ISI because significant CIR taps at other delay times may remain but the cost function is insensitive to them. 
A more recent experiment in Ref.~\cite{zhou2022multipath_Mitigation_RIS} targeted RIS-based multipath mitigation, albeit in a simple two-path propagation scenario that strongly differs from the rich scattering that WNoCs are confronted with.
Meanwhile, various model-based studies of CIR equalization using a RIS appeared~\cite{zhang2021spatial,arslan2021over,PhysFad,hugo_eucap}, of which Refs.~\cite{PhysFad,hugo_eucap} are based on a physics-compliant model (but describing an abstract 2D setup) that respects causality and correctly captures proximity-induced and reverberation-induced mutual coupling between the primary wireless entities (antennas and RIS elements). Since ISI mitigation requires that the optimized CIR has a single thin dominant tap, the optimization objective in Refs.~\cite{PhysFad,hugo_eucap} was formulated as the minimization of the ratio between total signal intensity and signal intensity within an interval $\Delta t$ centered on a delay time $\tau_0$, yielding the following cost function:
\begin{equation}
\mathcal{C}(\tau_0,\mathbf{c}) = \frac{\int_0^\infty h^2(t,\mathbf{c}) \mathrm{d}t}{\int_{\tau_0-\Delta t/2}^{\tau_0+\Delta t/2} h^2(t,\mathbf{c}) \mathrm{d}t}.
\label{eq:CF}
\end{equation}
In the absence of any ISI, e.g., in free space, $\mathcal{C} \rightarrow 1$ (unity is the lower bound for the values that $\mathcal{C}$ can take).
The success of ISI mitigation pivotally depends on the choice of the values for $\Delta t$ and $\tau_0$. With respect to $\Delta t$, the signal bandwidth $B$ imposes a lower bound $1/B$ on how thin temporal CIR features can become. Since the optimized CIR tap is desired to be as thin as possible, the best choice is hence $\Delta t = 1/B$~\cite{hugo_eucap}. 

But what value should be chosen for $\tau_0$? To answer this question, we now systematically evaluate how the achievable $\mathcal{C}$ depends on $\tau_0$ for the three wireless links of interest in our WNoC case study. The identification of the optimal RIS configuration (out of the $2^{64}$ possible ones) for a given choice of wireless link and $\tau_0$ cannot be performed in a closed analytical form because of the non-linear dependence of $h(t,\mathbf{c})$ on $\mathbf{c}$. Therefore, we perform an iterative optimization of the RIS configuration using Algorithm~\ref{algo1} for each considered wireless link and each considered value of $\tau_0$. Such a systematic investigation is enabled by our compact physical model and would be prohibitively expensive if one full-wave simulation at every iteration was required to evaluate $\mathcal{C}$. In fact, to efficiently implement Algorithm~\ref{algo1}, we do not even need to perform the evaluation of $H(f,\mathbf{c})$ as detailed in Sec.~\ref{sec_Theory} from scratch in every iteration. Instead, we update a previous channel evaluation to account for a new RIS configuration, akin to the method proposed in Sec.~IV.A of Ref.~\cite{prod2023efficient} based on the Woodbury identity in the context of a tunable-polarizability physics-compliant model.

\begin{algorithm}[h]
Evaluate the costs $\left\{\mathcal{C}\right\}_{\mathrm{init}}$ for $50$ random RIS configurations $\left\{\mathbf{c}\right\}_{\mathrm{init}}$.\\
Select $\mathcal{C}_{\rm{curr}}$ as the minimum cost from $\left\{\mathcal{C}\right\}_{\mathrm{init}}$ and $\mathbf{c}_{\mathrm{curr}}$ from $\left\{\mathbf{c}\right\}_{\mathrm{init}}$ at the same index such as: $\mathcal{C}_{\mathrm{curr}} = \mathcal{C}\left(\mathbf{c}_{\mathrm{curr}}\right) = \mathrm{min}\left(\left\{\mathcal{C}\right\}_{\mathrm{init}}\right)$.\\
$k \gets 0$\\
$i \gets 1$\\
\While{$k < N_\mathrm{S}$}{
    $i \gets i+1$\\
    $\mathbf{c}^\prime \gets \mathbf{c}_\mathrm{curr}$  with the $\mathrm{mod}(i,N_{\rm S})$th bit flipped.\\
    $\mathcal{C}_{\rm temp} \gets \mathcal{C}\left(\mathbf{c}^\prime\right)$.\\
    \eIf{
    $\mathcal{C}_{\rm temp} < \mathcal{C}_{\rm curr}$
    } {
    $\mathbf{c}_\mathrm{curr} \gets \mathbf{c}^\prime$ \\
    $\mathcal{C}_\mathrm{curr} \gets \mathcal{C}_\mathrm{temp}$ \\
    $k \gets 0$
    } {$k \gets k+1$}
    }
\KwOut{Optimized RIS configuration $\mathbf{c}_{\rm curr}$ and/or corresponding cost $\mathcal{C}_\mathrm{curr}$.}
\caption{Binary RIS optimization for OTA equalization of single CIR for given delay time $\tau_0$.}
\label{algo1}
\end{algorithm}

\begin{figure}[h]
\centering
\includegraphics[width=\linewidth]{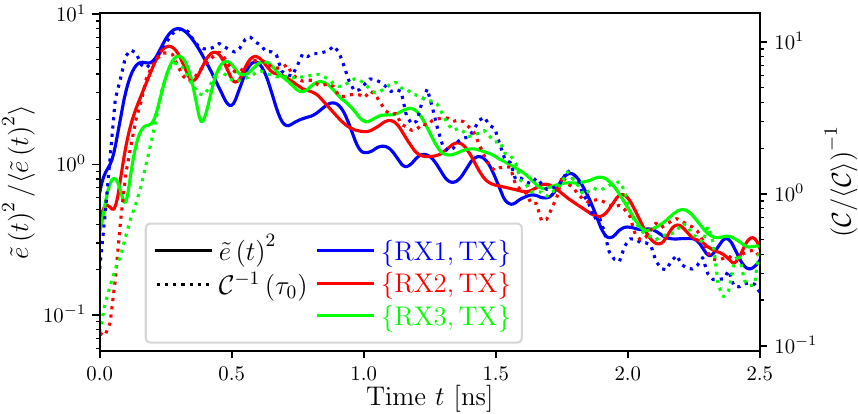}
\caption{Systematic investigation of achievable $\mathcal{C}$ as a function of $\tau_0$ for the three considered wireless on-chip channels (dashed lines, color-coded as in Fig.~\ref{fig_in_situ_charac}, using right hand vertical axis). Moreover, the corresponding average squared CIR envelopes are shown as a function of time (continuous lines, same color codes, using left hand vertical axis).}
\label{fig_results1}
\end{figure}

The dependence of the achievable cost function $\mathcal{C}$ on the choice of delay time $\tau_0$ is displayed by the dashed lines in Fig.~\ref{fig_results1}. The optimal choice of delay time for a given wireless link yields the smallest possible value of $\mathcal{C}$ (i.e., largest possible value of $\mathcal{C}^{-1}$) and is seen in Fig.~\ref{fig_results1} to be different for each considered wireless link. We hypothesize that the optimal choice of delay time strongly correlates with setting-specific features of the CIR: multi-bounce paths that were significantly impacted by the RIS play an important role at some delay times but not at others. At delay times at which paths that are unaffected by the RIS dominate, it is impossible to configure the RIS to create a good destructive interference. The CIR will tend to have a tap at a delay time with a peak of the average CIR, and if this tap is not within the $\Delta t$ interval around $\tau_0$, it will deteriorate the value of $\mathcal{C}$. Therefore, we hypothesize that $\tau_0$ should be chosen to correspond to the maximum of the average CIR envelope in order to achieve the smallest possible value of $\mathcal{C}$. 
To confirm our hypothesis, we evaluate the average CIR envelope:
\begin{equation}
    \Tilde{e}(t) = \langle e(t,\mathbf{c}) \rangle_\mathbf{c},
\end{equation}
where the average is taken over random RIS configurations. The average squared CIR envelopes are shown in Fig.~\ref{fig_results1} as continuous lines. As hypothesized, their maxima occur at the same delay times as the maxima of $\mathcal{C}$. Therefore, we have identified a simple procedure to determine the best choice of $\tau_0$: evaluating the CIR for a set of random RIS configurations and determining the delay time at which $\Tilde{e}(t) $ has its maximum. We use this insight in the subsequent section for simultaneous multi-link OTA equalization.

\begin{figure*}
\centering
\includegraphics[width=\linewidth]{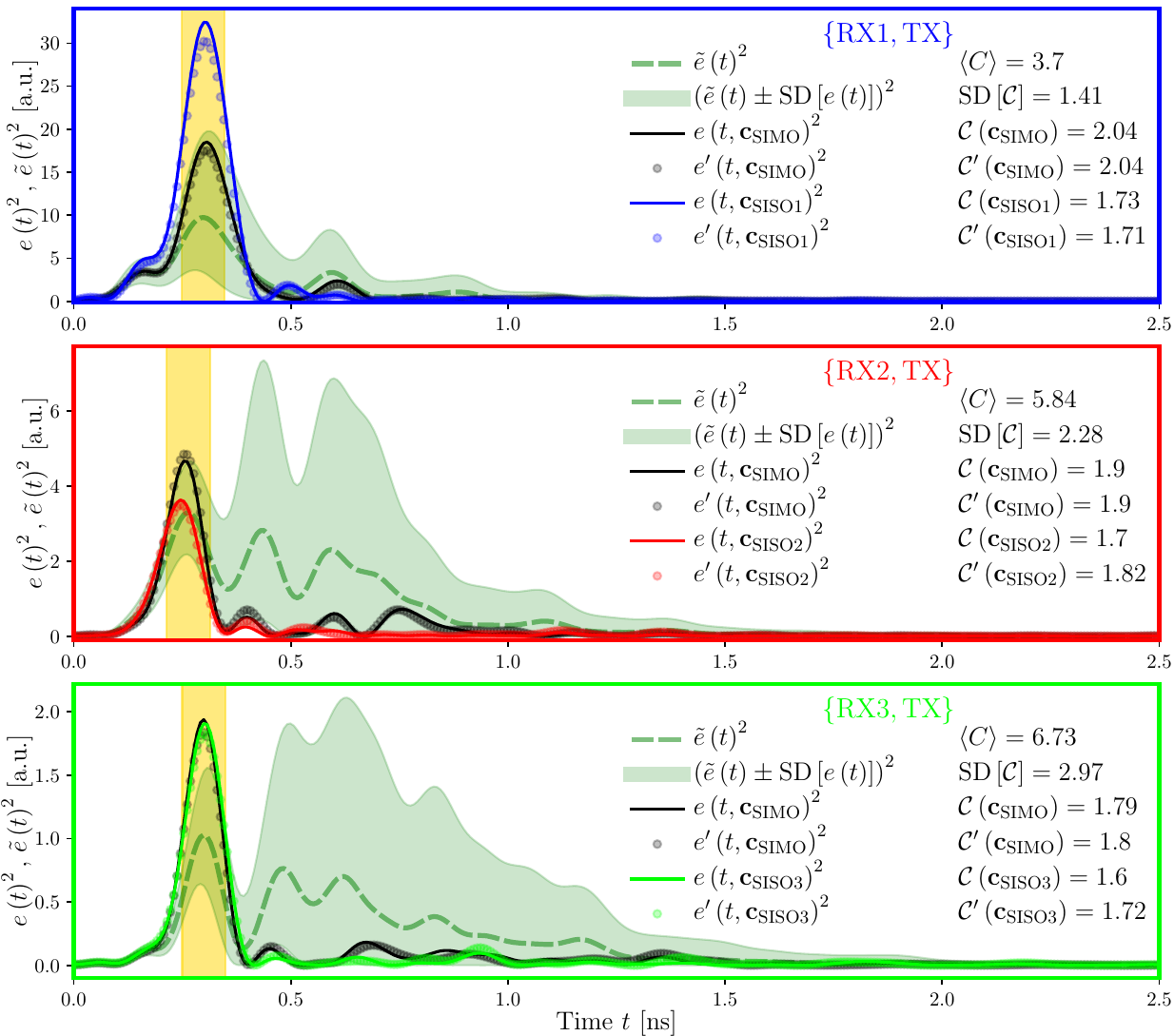}
\caption{RIS-optimized CIRs for OTA equalization. For each of the three considered on-chip wireless links (color-coded boxes, see Fig.~\ref{fig_in_situ_charac}a), the average CIR envelope and its standard deviation are shown as dashed line and shaded area, respectively. In addition, the optimal choices of $\Delta t$ and $\tau_0$ are indicated by the yellow shaded box. 
Furthermore, CIR envelopes are shown that correspond to the model's prediction (continuous lines, $e(t)$) and ground truth (dots, $e'(t)$) for the RIS configurations optimized for operation with SISO (colored) or SIMO (black). 
The insets indicate the corresponding values of the cost function based on the model's prediction ($\mathcal{C}$) and the corresponding ground truth values ($\mathcal{C}'$). The corresponding optimized RIS configurations are shown in Fig.~\ref{fig_results3}.}
\label{fig_results2}
\end{figure*}

For each of the three considered wireless links, the average CIR and its standard deviation are shown as dashed line and shaded area, respectively, in Fig.~\ref{fig_results2}. Moreover, the optimal time interval choice with $\tau_0$ corresponding to the peak of the average CIR envelope and $\Delta t = 1/B$ is highlighted in yellow. In addition, in each case, an optimized CIR for this specific link is shown in terms of the model's prediction (continuous colored line) and the corresponding ground truth (colored dots). The contribution to mitigating ISI is significant in all three cases: compared to the values of $\mathcal{C}$ for random RIS configurations ($3.7\pm1.4$, $5.8\pm2.3$, $6.7\pm3.0$), the optimized CIRs reach significantly lower values (predicted: $1.7$, $1.7$, $1.6$; ground truth: $1.7$, $1.8$, $1.7$) in all three considered cases. It is also apparent upon visual inspection that only one significant CIR tap remains for each of the optimized CIRs. We have run this optimization multiple times; different runs tend to yield different optimized configurations but they tend to produce similar values of $\mathcal{C}$. Moreover, we determined that the fraction of RIS elements configured to be ON in the optimized configurations is $0.53$. Being very close to 0.5, this fraction confirms that the investigated OTA equalization technique relies on intricate tailored interference effects rather than absorption (otherwise, all RIS elements would be in the ON configuration that absorbs more than the OFF configuration -- see Fig.~\ref{fig_ris_element}). This is also visually confirmed in Fig.~\ref{fig_results3}a-c by the three example configurations corresponding to the optimized CIRs shown in Fig.~\ref{fig_results2} for the SISO cases; Fig.~\ref{fig_results3}a-c further shows that the optimizations do not appear to yield intuitively interpretable patterns for the optimized RIS configuration, as expected due to the rich scattering.

\section{Multi-Link OTA Equalization \\for Broadcasting}
\label{sec_multi}

So far, we have only discussed the OTA equalization of a single WNoC CIR (i.e., single-input single-output, SISO) but some of the key promises held by WNoCs relate to broadcast scenarios in which one transmitter shares information with multiple receivers simultaneously (i.e., single-input multiple-output, SIMO). Therefore, we now consider the scenario in which the transmitting antenna seeks to simultaneously broadcast information to all three receiving antennas, meaning that all three CIRs should be simultaneously equalized OTA by a single optimized RIS configuration. Having identified in the previous Sec.~\ref{syststud} how to chose the value of $\tau_0$ for a given CIR, we now identify the optimal value of $\tau_0$ for each of the three considered links. This allows us to define a combined cost function: for a given RIS configuration $\mathbf{c}$, first, we evaluate the cost function of each individual link using Eq.~(\ref{eq:CF}) based on the optimal value of $\tau_0$ for that link; second, we evaluate a combined cost function $\overline{\mathcal{C}}$ that combines the individual cost functions in a root-mean-square fashion:
\begin{equation}
\overline{\mathcal{C}}\left(\mathbf{c}\right)=\sqrt{\langle\mathcal{C}^2_{i,j}\left(\mathbf{c},\tau_{0,i,j}\right)\rangle},
\label{eq:CF2}
\end{equation}
where the indices $i,j$ identify the different considered wireless on-chip links. We then use the same iterative procedure as in Algorithm~\ref{algo1} to optimize a RIS configuration for this multi-link SIMO case. The detailed procedure for the SIMO case is summarized in Algorithm~\ref{algo2}.

\begin{algorithm}[h]
Evaluate $\tau_{0,i,j}$ for $i\in \mathcal{R}$ and $j \in \mathcal{T}$.\\
Define $\overline{\mathcal{C}}\left(\mathbf{c}\right)=\sqrt{\langle\mathcal{C}^2_{i,j}\left(\mathbf{c},\tau_{0,i,j}\right)\rangle}$.\\
Apply Algorithm~\ref{algo1} using $\overline{\mathcal{C}}$ as the cost function.\\
\KwOut{Optimized RIS configuration $\mathbf{c}_{\rm curr}$ and/or corresponding cost $\mathcal{C}_\mathrm{curr}$.}
\caption{Binary RIS Optimization for Simultaneous OTA Equalization of Multiple CIRs}
\label{algo2}
\end{algorithm}

\begin{figure}
\centering
\includegraphics[width=\linewidth]{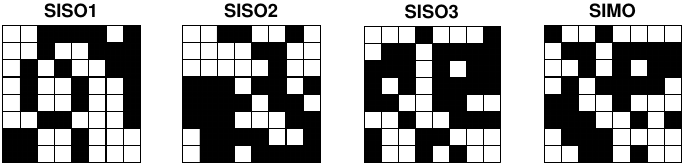}
\caption{Optimized RIS configurations corresponding to the results displayed in Fig.~\ref{fig_results2}.}
\label{fig_results3}
\end{figure}

The value of $\overline{\mathcal{C}}$ for random RIS configurations for the three considered wireless on-chip links is $5.6\pm 1.8$. CIRs corresponding to the optimized RIS configuration obtained with one realization of Algorithm~\ref{algo2} are shown in black in Fig.~\ref{fig_results2} (model prediction: black continuous line; ground truth: black dots). This optimized RIS configuration achieves $\overline{\mathcal{C}}=1.9$, and the values of $\mathcal{C}$ for each wireless link are $2.0$, $1.9$ and $1.8$ (model predictions and ground truths are identical). 
Of course, the optimization objective is more demanding in this SIMO case, as compared to the previous SISO cases, but nonetheless a significant contribution to ISI mitigation is apparent. Again, upon visual inspection we see that the optimized CIRs only have one dominant tap, as required for ISI-free OOK communications. The results from Fig.~\ref{fig_results2} confirm the suitability of the proposed RIS-based OTA equalization technique for broadcast WNoC scenarios. 
We have run Algorithm~\ref{algo2} multiple times and found that for the optimized RIS configurations the ratio of RIS elements in the ON state is 0.51, again extremely close to 0.5 (see also the displayed example optimized RIS configuration for SIMO in Fig.~\ref{fig_results3}d). We observe more significant fluctuations in the values of $\mathcal{C}$ with the optimized RIS configurations for Algorithm~\ref{algo2} than we did for Algorithm~\ref{algo1}, suggesting that future work can refine the combined cost function from Eq.~(\ref{eq:CF2}) to yield a more stable performance.

\section{Conclusion}\label{sec_Conclusion}

To summarize, we have introduced and validated a physics-compliant model for RIS-parametrized WNoC channels that can be calibrated with a single full-wave simulation. Based on our model, we systematically studied the choice of delay time for the optimized CIR's peak, finding that it should coincide with the time at which the average CIR envelope (averaged over random RIS configuration) has its maximum. Then, we tackled a broadcast scenario and showed that multi-link RIS-based OTA equalization is feasible in WNoCs. Parametrizing a WNoC with an RIS is hence a promising route to mitigating ISI without loss of signal strength nor the need for a resource-demanding modulation scheme.

Incidentally, given the scale invariance of Maxwell's equations, our approach to calibrating a physics-compliant model of a RIS-parametrized radio environment can also be transposed to the analysis of ``smart'' indoor radio environments whenever such an analysis is based on full-wave simulations. 

Looking forward, on the one hand, we envision the transposition of the present study to the THz regime where more bandwidth is readily available than in the here considered millimeter-wave regime: a given bandwidth (WNoCs target $10-15 \ \mathrm{GHz})$ constitutes a much smaller fractional bandwidth~\cite{tapie2023toward}. Various THz RIS designs operating in reflection have already been proposed and experimentally prototyped in the literature~\cite{lynch2019coded,wu2020liquid,chen2022electrically,monroe2022electronic,lan2023real,liu2023toward} and could be adapted to the on-chip scenario. Signal generation becomes increasingly challenging at such high frequencies but could be implemented by distributing a centrally generated signal.
On the other hand, the tools introduced in the present paper enable a systematic study of OTA analog computing in WNoCs. In fact, the presented OTA equalization can already be interpreted as a confluence of analog computing and communications. Given the scale invariance of Maxwell's equations, concepts for OTA analog computing for matrix multiplication~\cite{del2018leveraging}, signal differentiation~\cite{sol2022meta} or physical wave-based neural networks~\cite{momeni2023backpropagation} that have already been prototyped in experiments at the indoor scale can be transposed to the chip scale. Finally, we note that the concept underlying the RIS-parametrized WNoCs, namely the massive parametrization of a complex scattering system to tune its scattering properties in situ, in principle also applies to waveguide networks and thus may also prove useful for certain implementations of guided NoCs.

\appendices

\section{Derivation of Eq.~(\ref{Zaug})}\label{AppendixA}

The impedance matrix $\mathbf{\mathring{Z}}\in\mathbb{C}^{M \times  M}$ of an $M$-port system relates port voltages $\mathbf{\mathring{V}} \in\mathbb{C}^{M \times  1} $ to port currents $\mathbf{\mathring{J}}\in\mathbb{C}^{M \times  1}$:
\begin{equation}
    \mathbf{\mathring{V}} = \mathbf{\mathring{Z}} \mathbf{\mathring{J}}.
\end{equation}
If the $m$th port is terminated with a load impedance $\mathring{\eta}_m$, this imposes the relation $\mathring{V}_m = -\mathring{\eta}_m \mathring{J}_m$ ($\mathring{J}_m$ is defined as the current going into the port). 

In our problem, $N_\mathrm{S}$ ports of the $N$ ports defining $\mathbf{\hat{Z}}$ are terminated by the load impedances summarized in $\mathbf{c}$. We partition $\mathbf{\hat{V}} \in\mathbb{C}^{N \times  1} $, $\mathbf{\hat{J}} \in\mathbb{C}^{N \times  1} $ and $\mathbf{\hat{Z}}$ according to the sets $\mathcal{A}$ and $\mathcal{S}$:
\begin{equation}
    \mathbf{\hat{V}} = \begin{bmatrix} 
	 \mathbf{\hat{V}}_\mathcal{A}   \\	 \mathbf{\hat{V}}_\mathcal{S}
 \end{bmatrix}, \  \mathbf{\hat{J}} = \begin{bmatrix} 
	 \mathbf{\hat{J}}_\mathcal{A}   \\	 \mathbf{\hat{J}}_\mathcal{S}
 \end{bmatrix}, \  \mathbf{\hat{Z}} = \begin{bmatrix} 
	 \mathbf{\hat{Z}}_\mathcal{AA}  &   \mathbf{\hat{Z}}_\mathcal{AS} \\	 \mathbf{\hat{Z}}_\mathcal{SA}  & \mathbf{\hat{V}}_\mathcal{SS}
 \end{bmatrix},
\end{equation}
and use
\begin{equation}
    \mathbf{\hat{V}}_\mathcal{S} = -\mathbf{\Phi} \mathbf{\hat{J}}_\mathcal{S}
\end{equation}
to obtain
\begin{equation}
     \begin{bmatrix} 
	 \mathbf{\hat{V}}_\mathcal{A}   \\	 \mathbf{0}_\mathcal{S}
 \end{bmatrix} + \begin{bmatrix} 
	 \mathbf{0}_\mathcal{AA} & \mathbf{0}_\mathcal{AS}   \\	 \mathbf{0}_\mathcal{SA} & -\mathbf{\Phi}
 \end{bmatrix} \begin{bmatrix} 
	 \mathbf{\hat{J}}_\mathcal{A}   \\	 \mathbf{\hat{J}}_\mathcal{S}
 \end{bmatrix}= \begin{bmatrix} 
	 \mathbf{\hat{Z}}_\mathcal{AA}  &   \mathbf{\hat{Z}}_\mathcal{AS} \\	 \mathbf{\hat{Z}}_\mathcal{SA}  & \mathbf{\hat{Z}}_\mathcal{SS}
 \end{bmatrix} \begin{bmatrix} 
	 \mathbf{\hat{J}}_\mathcal{A}   \\	 \mathbf{\hat{J}}_\mathcal{S}
 \end{bmatrix}.
 \label{eqX}
\end{equation}
Rearranging Eq.~(\ref{eqX}), we obtain
\begin{equation}
    \begin{bmatrix} 
	 \mathbf{\hat{V}}_\mathcal{A}   \\	 \mathbf{0}_\mathcal{S}
 \end{bmatrix} = \begin{bmatrix} 
	 \mathbf{\hat{Z}}_\mathcal{AA}  &   \mathbf{\hat{Z}}_\mathcal{AS} \\	 \mathbf{\hat{Z}}_\mathcal{SA}  & \mathbf{\hat{Z}}_\mathcal{SS}+\mathbf{\Phi}
 \end{bmatrix} \begin{bmatrix} 
	 \mathbf{\hat{J}}_\mathcal{A}   \\	 \mathbf{\hat{J}}_\mathcal{S}
 \end{bmatrix}
\end{equation}
which implies
\begin{subequations}
\begin{equation}
     \mathbf{\hat{V}}_\mathcal{A} = \mathbf{\hat{Z}}_\mathcal{AA}\mathbf{\hat{J}}_\mathcal{A} + \mathbf{\hat{Z}}_\mathcal{AS}\mathbf{\hat{J}}_\mathcal{S}.
     \label{eqyyy}
\end{equation}
\begin{equation}
     \mathbf{0}_\mathcal{S} = \mathbf{\hat{Z}}_\mathcal{SA}\mathbf{\hat{J}}_\mathcal{A} + \left(\mathbf{\hat{Z}}_\mathcal{AS} + \mathbf{\Phi}\right)\mathbf{\hat{J}}_\mathcal{S}.
     \label{eqyy}
\end{equation}
\end{subequations}
Solving Eq.~(\ref{eqyy}) for $\mathbf{{J}}_\mathcal{S}$ yields
\begin{equation}
    \mathbf{\hat{J}}_\mathcal{S} = - \left(\mathbf{\hat{Z}}_\mathcal{AS} + \mathbf{\Phi}\right)^{-1}\mathbf{\hat{Z}}_\mathcal{SA}\mathbf{\hat{J}}_\mathcal{A},
    \label{eqyyyy}
\end{equation}
and inserting Eq.~(\ref{eqyyyy}) into Eq.~(\ref{eqyyy}) yields
\begin{equation}
    \mathbf{\hat{V}}_\mathcal{A} = \mathbf{\hat{Z}}_\mathcal{AA}\mathbf{\hat{J}}_\mathcal{A} - \mathbf{\hat{Z}}_\mathcal{AS}\left(\mathbf{\hat{Z}}_\mathcal{AS} + \mathbf{\Phi}\right)^{-1}\mathbf{\hat{Z}}_\mathcal{SA}\mathbf{\hat{J}}_\mathcal{A}.
    \label{sfdsdf}
\end{equation}
Since $\mathbf{\hat{V}}_\mathcal{A} = \mathbf{Z}\mathbf{\hat{J}}_\mathcal{A}$ in our problem, Eq.~(\ref{Zaug}) follows by inspection of Eq.~(\ref{sfdsdf}).

To avoid confusion regarding sign conventions, there are two simple sanity checks that can be performed to confirm the correctness of Eq.~(\ref{Zaug}):
\begin{enumerate}
    \item If all $N_\mathrm{S}$ ports with indices included in $\mathcal{S}$ are terminated with matched loads (i.e., $\mathbf{\Phi} = Z_0 \mathbf{I}_{N_\mathrm{S}}$), one should find that $\mathbf{S} = \mathbf{\hat{S}}_\mathcal{AA}$, where $\mathbf{\hat{S}} = \left( \mathbf{\hat{Z}} + Z_0 \mathbf{I}_{N}\right)^{-1} \left( \mathbf{\hat{Z}} - Z_0 \mathbf{I}_{N}\right)$.
    \item If all $N_\mathrm{S}$ ports with indices included in $\mathcal{S}$ are terminated with open circuits (i.e., $\mathbf{\Phi} = \eta \mathbf{I}_{N_\mathrm{S}}$ with $\eta \rightarrow \infty$), one should find that $\mathbf{Z} = \mathbf{\hat{Z}}_\mathcal{AA}$.
\end{enumerate}

\bibliographystyle{IEEEtran}

\begin{thebibliography}{10}
\providecommand{\url}[1]{#1}
\csname url@samestyle\endcsname
\providecommand{\newblock}{\relax}
\providecommand{\bibinfo}[2]{#2}
\providecommand{\BIBentrySTDinterwordspacing}{\spaceskip=0pt\relax}
\providecommand{\BIBentryALTinterwordstretchfactor}{4}
\providecommand{\BIBentryALTinterwordspacing}{\spaceskip=\fontdimen2\font plus
\BIBentryALTinterwordstretchfactor\fontdimen3\font minus \fontdimen4\font\relax}
\providecommand{\BIBforeignlanguage}[2]{{%
\expandafter\ifx\csname l@#1\endcsname\relax
\typeout{** WARNING: IEEEtran.bst: No hyphenation pattern has been}%
\typeout{** loaded for the language `#1'. Using the pattern for}%
\typeout{** the default language instead.}%
\else
\language=\csname l@#1\endcsname
\fi
#2}}
\providecommand{\BIBdecl}{\relax}
\BIBdecl

\bibitem{nychis2012chip}
G.~P. Nychis, C.~Fallin, T.~Moscibroda, O.~Mutlu, and S.~Seshan, ``On-chip networks from a networking perspective: Congestion and scalability in many-core interconnects,'' \emph{ACM SIGCOMM Comput. Commun. Rev.}, vol.~42, no.~4, pp. 407--418, 2012.

\bibitem{karkar2016survey}
A.~Karkar, T.~Mak, K.-F. Tong, and A.~Yakovlev, ``A survey of emerging interconnects for on-chip efficient multicast and broadcast in many-cores,'' \emph{IEEE Circuits Syst. Mag.}, vol.~16, no.~1, pp. 58--72, 2016.

\bibitem{wentzlaff2007chip}
D.~Wentzlaff, P.~Griffin, H.~Hoffmann, L.~Bao, B.~Edwards, C.~Ramey, M.~Mattina, C.-C. Miao, J.~F. Brown~III, and A.~Agarwal, ``On-chip interconnection architecture of the tile processor,'' \emph{IEEE Micro}, vol.~27, no.~5, pp. 15--31, 2007.

\bibitem{marculescu2008outstanding}
R.~Marculescu, U.~Y. Ogras, L.-S. Peh, N.~E. Jerger, and Y.~Hoskote, ``Outstanding research problems in noc design: system, microarchitecture, and circuit perspectives,'' \emph{IEEE Trans. Comput.-Aided Des.}, vol.~28, no.~1, pp. 3--21, 2008.

\bibitem{todri}
A.~Todri-Sanial, R.~Ramos, H.~Okuno, J.~Dijon, A.~Dhavamani, M.~Widlicenus, K.~Lilienthal, B.~Uhlig, T.~Sadi, V.~Georgiev, A.~Asenov, S.~Amoroso, A.~Pender, A.~Brown, C.~Millar, F.~Motzfeld, B.~Gotsmann, J.~Liang, G.~Goncalves, N.~Rupesinghe, and K.~Teo, ``A survey of carbon nanotube interconnects for energy efficient integrated circuits,'' \emph{IEEE Circuits Syst. Mag.}, vol.~17, no.~2, pp. 47--62, 2017.

\bibitem{franques2021widir}
A.~Franques, A.~Kokolis, S.~Abadal, V.~Fernando, S.~Misailovic, and J.~Torrellas, ``Widir: A wireless-enabled directory cache coherence protocol,'' \emph{Proc. HPCA}, pp. 304--317, 2021.

\bibitem{miller2009device}
D.~A. Miller, ``Device requirements for optical interconnects to silicon chips,'' \emph{Proc. IEEE}, vol.~97, no.~7, pp. 1166--1185, 2009.

\bibitem{dionne2010silicon}
J.~A. Dionne, L.~A. Sweatlock, M.~T. Sheldon, A.~P. Alivisatos, and H.~A. Atwater, ``Silicon-based plasmonics for on-chip photonics,'' \emph{IEEE J. Sel. Top. Quantum Electron.}, vol.~16, no.~1, pp. 295--306, 2010.

\bibitem{sun2015single}
C.~Sun, M.~T. Wade, Y.~Lee, J.~S. Orcutt, L.~Alloatti, M.~S. Georgas, A.~S. Waterman, J.~M. Shainline, R.~R. Avizienis, S.~Lin \emph{et~al.}, ``Single-chip microprocessor that communicates directly using light,'' \emph{Nature}, vol. 528, no. 7583, pp. 534--538, 2015.

\bibitem{wade2020teraphy}
M.~Wade, E.~Anderson, S.~Ardalan, P.~Bhargava, S.~Buchbinder, M.~L. Davenport, J.~Fini, H.~Lu, C.~Li, R.~Meade \emph{et~al.}, ``Teraphy: a chiplet technology for low-power, high-bandwidth in-package optical i/o,'' \emph{IEEE Micro}, vol.~40, no.~2, pp. 63--71, 2020.

\bibitem{chang2001rf}
M.~F. Chang, V.~P. Roychowdhury, L.~Zhang, H.~Shin, and Y.~Qian, ``Rf/wireless interconnect for inter-and intra-chip communications,'' \emph{Proc. IEEE}, vol.~89, no.~4, pp. 456--466, 2001.

\bibitem{ganguly2010scalable}
A.~Ganguly, K.~Chang, S.~Deb, P.~P. Pande, B.~Belzer, and C.~Teuscher, ``Scalable hybrid wireless network-on-chip architectures for multicore systems,'' \emph{IEEE Trans. Comput.}, vol.~60, no.~10, pp. 1485--1502, 2010.

\bibitem{matolak2012wireless}
D.~W. Matolak, A.~Kodi, S.~Kaya, D.~Ditomaso, S.~Laha, and W.~Rayess, ``Wireless networks-on-chips: architecture, wireless channel, and devices,'' \emph{IEEE Wirel. Commun.}, vol.~19, no.~5, pp. 58--65, 2012.

\bibitem{deb}
S.~Deb, A.~Ganguly, P.~P. Pande, B.~Belzer, and D.~Heo, ``Wireless noc as interconnection backbone for multicore chips: Promises and challenges,'' \emph{IEEE J. Emerg. Sel. Top. Circuits Syst.}, vol.~2, no.~2, pp. 228--239, 2012.

\bibitem{6814853}
X.~Yu, J.~Baylon, P.~Wettin, D.~Heo, P.~P. Pande, and S.~Mirabbasi, ``Architecture and design of multichannel millimeter-wave wireless {NoC},'' \emph{IEEE Des. Test}, vol.~31, no.~6, pp. 19--28, 2014.

\bibitem{shamim2016wireless}
M.~S. Shamim, N.~Mansoor, R.~S. Narde, V.~Kothandapani, A.~Ganguly, and J.~Venkataraman, ``A wireless interconnection framework for seamless inter and intra-chip communication in multichip systems,'' \emph{IEEE Trans. Comput.}, vol.~66, no.~3, pp. 389--402, 2016.

\bibitem{abadal2022graphene}
S.~Abadal, R.~Guirado, H.~Taghvaee, A.~Jain, E.~P. de~Santana, P.~H. Bolivar, M.~Saeed, R.~Negra, Z.~Wang, K.-T. Wang \emph{et~al.}, ``Graphene-based wireless agile interconnects for massive heterogeneous multi-chip processors,'' \emph{IEEE Wirel. Commun.}, 2022.

\bibitem{calo2022reconfigurable}
G.~Cal{\`o}, L.~Gabriele, G.~Bellanca, J.~Nanni, M.~Barbiroli, F.~Fuschini, V.~Tralli, D.~Bertozzi, G.~Serafino, and V.~Petruzzelli, ``Reconfigurable optical wireless switches for on-chip interconnection,'' \emph{IEEE J. Quantum Electron.}, 2022.

\bibitem{laha2014new}
S.~Laha, S.~Kaya, D.~W. Matolak, W.~Rayess, D.~DiTomaso, and A.~Kodi, ``A new frontier in ultralow power wireless links: Network-on-chip and chip-to-chip interconnects,'' \emph{IEEE Trans. Comput.-Aided Des.}, vol.~34, no.~2, pp. 186--198, 2014.

\bibitem{markish2015chip}
O.~Markish, O.~Katz, B.~Sheinman, D.~Corcos, and D.~Elad, ``On-chip millimeter wave antennas and transceivers,'' in \emph{Proc. NoCS}, 2015, pp. 1--7.

\bibitem{abadal2019wave}
S.~Abadal, C.~Han, and J.~M. Jornet, ``Wave propagation and channel modeling in chip-scale wireless communications: A survey from millimeter-wave to terahertz and optics,'' \emph{IEEE Access}, vol.~8, pp. 278--293, 2019.

\bibitem{matolak2013channel}
D.~W. Matolak, S.~Kaya, and A.~Kodi, ``Channel modeling for wireless networks-on-chips,'' \emph{IEEE Commun. Mag.}, vol.~51, no.~6, pp. 180--186, 2013.

\bibitem{8512165}
J.~O. Sosa, O.~Sentieys, and C.~Roland, ``A diversity scheme to enhance the reliability of wireless {NoC} in multipath channel environment,'' \emph{Proc. NoCS}, pp. 1--8, 2018.

\bibitem{fatimaTR}
A.~Bandara, F.~Rodr\'{\i}guez-Gal\'{a}n, E.~P. de~Santana, P.~H. Bol\'{\i}var, E.~Alarc\'{o}n, and S.~Abadal, ``Exploration of time reversal for wireless communications within computing packages,'' \emph{Proc. ACM NanoCom}, p. 14–20, 2023.

\bibitem{timoneda2020engineer}
X.~Timoneda, S.~Abadal, A.~Franques, D.~Manessis, J.~Zhou, J.~Torrellas, E.~Alarc{\'o}n, and A.~Cabellos-Aparicio, ``Engineer the channel and adapt to it: Enabling wireless intra-chip communication,'' \emph{IEEE Trans. Commun.}, vol.~68, no.~5, pp. 3247--3258, 2020.

\bibitem{treguer2023broadband}
B.~Treguer, T.~Le~Gouguec, P.-M. Martin, R.~Allanic, and C.~Quendo, ``Broadband silicon controlled channel for wireless network-on-chip at 60 {GHz},'' \emph{IEEE Access}, 2023.

\bibitem{imani2021smart}
M.~F.~Imani, S.~Abadal, and P.~del Hougne, ``Metasurface-programmable wireless network-on-chip,'' \emph{Adv. Sci.}, vol.~9, no.~26, p. 2201458, 2022.

\bibitem{rabault2023tacit}
A.~Rabault \emph{et~al.}, ``On the tacit linearity assumption in common cascaded models of {RIS}-parametrized wireless channels,'' \emph{arXiv:2302.04993}, 2023.

\bibitem{gradoni_EndtoEnd_2020}
G.~Gradoni and M.~Di~Renzo, ``End-to-end mutual coupling aware communication model for reconfigurable intelligent surfaces: An electromagnetic-compliant approach based on mutual impedances,'' \emph{IEEE Wirel. Commun. Lett.}, vol.~10, no.~5, pp. 938--942, 2021.

\bibitem{shen2021modeling}
S.~Shen, B.~Clerckx, and R.~Murch, ``Modeling and architecture design of reconfigurable intelligent surfaces using scattering parameter network analysis,'' \emph{IEEE Trans. Wirel. Commun.}, vol.~21, no.~2, pp. 1229--1243, 2021.

\bibitem{badheka2023accurate}
D.~Badheka, J.~Sapis, S.~R. Khosravirad, and H.~Viswanathan, ``Accurate modeling of intelligent reflecting surface for communication systems,'' \emph{IEEE Trans. Wirel. Commun.}, 2023.

\bibitem{akrout2023physically}
M.~Akrout, F.~Bellili, A.~Mezghani, and J.~A. Nossek, ``Physically consistent models for intelligent reflective surface-assisted communications under mutual coupling and element size constraint,'' \emph{arXiv:2302.11130}, 2023.

\bibitem{mursia2023saris}
P.~Mursia, S.~Phang, V.~Sciancalepore, G.~Gradoni, and M.~Di~Renzo, ``{SARIS}: Scattering aware reconfigurable intelligent surface model and optimization for complex propagation channels,'' \emph{IEEE Wirel. Commun. Lett.}, 2023.

\bibitem{PhysFad}
R.~Faqiri, C.~Saigre-Tardif, G.~C. Alexandropoulos, N.~Shlezinger, M.~F. Imani, and P.~del Hougne, ``{PhysFad}: Physics-based end-to-end channel modeling of {RIS}-parametrized environments with adjustable fading,'' \emph{IEEE Trans. Wirel. Commun.}, vol.~22, no.~1, pp. 580--595, 2023.

\bibitem{prod2023efficient}
H.~Prod'homme and P.~del Hougne, ``Efficient computation of physics-compliant channel realizations for (rich-scattering) {RIS}-parametrized radio environments,'' \emph{arXiv:2306.00244}, 2023.

\bibitem{sol2023experimentally}
J.~Sol, H.~Prod'Homme, L.~Le~Magoarou, and P.~del Hougne, ``Experimentally realized physical-model-based wave control in metasurface-programmable complex media,'' \emph{arXiv:2308.02349}, 2023.

\bibitem{hugo_eucap}
H.~Prod'homme, M.~F.~Imani, S.~Abadal, and P.~del Hougne, ``{RIS}-based over-the-air channel equalization in resource-constrained wireless networks,'' \emph{Proc. EuCAP}, 2024.

\bibitem{ivrlavc2010toward}
M.~T. Ivrla{\v{c}} and J.~A. Nossek, ``Toward a circuit theory of communication,'' \emph{IEEE Trans. Circuits Syst. I: Regul. Pap.}, vol.~57, no.~7, pp. 1663--1683, 2010.

\bibitem{zhang2007propagation}
Y.~P. Zhang, Z.~M. Chen, and M.~Sun, ``Propagation mechanisms of radio waves over intra-chip channels with integrated antennas: Frequency-domain measurements and time-domain analysis,'' \emph{IEEE Trans. Antennas Propag.}, vol.~55, no.~10, pp. 2900--2906, 2007.

\bibitem{konno2023generalised}
K.~Konno, S.~Terranova, Q.~Chen, and G.~Gradoni, ``Generalised impedance model of wireless links assisted by reconfigurable intelligent surfaces,'' \emph{arXiv:2306.03761}, 2023.

\bibitem{rayess2017antennas}
W.~Rayess, D.~W. Matolak, S.~Kaya, and A.~K. Kodi, ``Antennas and channel characteristics for wireless networks on chips,'' \emph{Wirel. Pers. Commun.}, vol.~95, pp. 5039--5056, 2017.

\bibitem{narde2019intra}
R.~S. Narde, J.~Venkataraman, A.~Ganguly, and I.~Puchades, ``Intra-and inter-chip transmission of millimeter-wave interconnects in noc-based multi-chip systems,'' \emph{IEEE Access}, vol.~7, pp. 112\,200--112\,215, 2019.

\bibitem{narde2020antenna}
------, ``Antenna arrays as millimeter-wave wireless interconnects in multichip systems,'' \emph{IEEE Antennas Wirel. Propag. Lett.}, vol.~19, no.~11, pp. 1973--1977, 2020.

\bibitem{macom}
\BIBentryALTinterwordspacing
MACOM. {MADP-000907-14020x}. [Online]. Available: \url{https://cdn.macom.com/datasheets/MADP-000907-14020x.pdf}
\BIBentrySTDinterwordspacing

\bibitem{del2016spatiotemporal}
P.~del Hougne, F.~Lemoult, M.~Fink, and G.~Lerosey, ``Spatiotemporal wave front shaping in a microwave cavity,'' \emph{Phys. Rev. Lett.}, vol. 117, no.~13, p. 134302, 2016.

\bibitem{zhou2022multipath_Mitigation_RIS}
R.~Zhou, X.~Chen, W.~Tang, X.~Li, S.~Jin, E.~Basar, Q.~Cheng, and T.~J. Cui, ``Modeling and measurements for multi-path mitigation with reconfigurable intelligent surfaces,'' \emph{Proc. EuCAP}, pp. 1--5, 2022.

\bibitem{zhang2021spatial}
H.~Zhang, L.~Song, Z.~Han, and H.~V. Poor, ``Spatial equalization before reception: reconfigurable intelligent surfaces for multi-path mitigation,'' \emph{Proc. ICASSP}, pp. 8062--8066, 2021.

\bibitem{arslan2021over}
E.~Arslan, I.~Yildirim, F.~Kilinc, and E.~Basar, ``Over-the-air equalization with reconfigurable intelligent surfaces,'' \emph{IET Commun.}, vol.~16, no.~13, pp. 1486--1497, 2022.

\bibitem{tapie2023toward}
J.~Tapie, M.~F.~Imani, and P.~del Hougne, ``Toward {THz RIS}-parametrized wireless networks-on-chip,'' in \emph{Proc. ACM NanoCom}, 2023, pp. 142--143.

\bibitem{lynch2019coded}
J.~J. Lynch, F.~G. Herrault, G.~L. Virbila, K.~S. Kona, D.~L. Hammon, D.~C. Regan, J.~C. Wong, Y.~Tang, E.~M. Prophet, P.~Naghibi \emph{et~al.}, ``Coded aperture subreflector array for high resolution radar imaging,'' \emph{Proc. SPIE}, vol. 10994, pp. 43--53, 2019.

\bibitem{wu2020liquid}
J.~Wu, Z.~Shen, S.~Ge, B.~Chen, Z.~Shen, T.~Wang, C.~Zhang, W.~Hu, K.~Fan, W.~Padilla, Y.~L. Lu, B.~Jun, J.~Chen, and P.~Wu, ``Liquid crystal programmable metasurface for terahertz beam steering,'' \emph{Appl. Phys. Lett.}, vol. 116, no.~13, 2020.

\bibitem{chen2022electrically}
B.~Chen, X.~Wang, W.~Li, C.~Li, Z.~Wang, H.~Guo, J.~Wu, K.~Fan, C.~Zhang, Y.~He, B.~Jin, J.~Chen, and P.~Wu, ``Electrically addressable integrated intelligent terahertz metasurface,'' \emph{Sci. Adv.}, vol.~8, no.~41, p. eadd1296, 2022.

\bibitem{monroe2022electronic}
N.~M. Monroe, G.~C. Dogiamis, R.~Stingel, P.~Myers, X.~Chen, and R.~Han, ``Electronic {THz} pencil beam forming and 2d steering for high angular-resolution operation: A 98$\times 98$-unit 265{GHz} cmos reflectarray with in-unit digital beam shaping and squint correction,'' \emph{Proc. ISSCC}, vol.~65, pp. 1--3, 2022.

\bibitem{lan2023real}
F.~Lan, L.~Wang, H.~Zeng, S.~Liang, T.~Song, W.~Liu, P.~Mazumder, Z.~Yang, Y.~Zhang, and D.~M. Mittleman, ``Real-time programmable metasurface for terahertz multifunctional wave front engineering,'' \emph{Light Sci. Appl.}, vol.~12, no.~1, p. 191, 2023.

\bibitem{liu2023toward}
Y.~Liu, Y.~Wang, X.~Fu, L.~Shi, F.~Yang, J.~Luo, Q.~Y. Zhou, Y.~Fu, Q.~Chen, J.~Y. Dai \emph{et~al.}, ``Toward sub-terahertz: Space-time coding metasurface transmitter for wideband wireless communications,'' \emph{Adv. Sci.}, p. 2304278, 2023.

\bibitem{del2018leveraging}
P.~del Hougne and G.~Lerosey, ``Leveraging chaos for wave-based analog computation: Demonstration with indoor wireless communication signals,'' \emph{Phys. Rev. X}, vol.~8, no.~4, p. 041037, 2018.

\bibitem{sol2022meta}
J.~Sol, D.~R. Smith, and P.~del Hougne, ``Meta-programmable analog differentiator,'' \emph{Nat. Commun.}, vol.~13, no.~1, pp. 1--10, 2022.

\bibitem{momeni2023backpropagation}
A.~Momeni, B.~Rahmani, M.~Mallejac, P.~del Hougne, and R.~Fleury, ``Backpropagation-free training of deep physical neural networks,'' \emph{arXiv:2304.11042}, 2023.

\end{thebibliography}


\end{document}